\documentclass{article}

\usepackage{PRIMEarxiv}

\usepackage{framed}

\usepackage{amssymb,amsmath}
\usepackage{pifont}

\usepackage{latexsym}

\usepackage{tabularx, multirow, makecell}
\newcolumntype{Y}{>{\centering\arraybackslash}X}

\usepackage{tablefootnote}
\usepackage{threeparttable}
\usepackage{colortbl}
\usepackage{caption}

\usepackage{url}

\usepackage{scalerel,stackengine,amsmath}

\usepackage{xcolor}
\usepackage{graphics}
\usepackage[hidelinks]{hyperref}

\usepackage[utf8]{inputenc} 
\usepackage[T1]{fontenc}    
\usepackage{hyperref}       
\usepackage{url}            
\usepackage{booktabs}       
\usepackage{amsfonts}       
\usepackage{nicefrac}       
\usepackage{microtype}      
\usepackage{lipsum}
\usepackage{fancyhdr}       
\graphicspath{{media/}}     

\title{Deep Learning-Based Prediction of Fractional Flow Reserve along the Coronary Artery
}

\author{Nils Hampe, Sanne G. M. van Velzen, Jean-Paul Aben, Carlos Collet, Ivana I\v sgum}

\begin{document}
\maketitle

\begin{abstract}
Functionally significant coronary artery disease (CAD) is caused by plaque buildup in the coronary arteries, potentially leading to a narrowing of the arterial lumen, i.e. coronary stenosis, that significantly obstructs the blood flow to the myocardium.
The current reference for establishing the presence of a functionally significant stenosis is invasive fractional flow reserve (FFR) measurement. 
To avoid invasive measurements, which are often shown unnecessary, non-invasive prediction of FFR from coronary CT angiography (CCTA) is an area of intensive research.
Increasingly, machine learning approaches, characterized by fast inference, are developed for the task. However, these methods predict a single FFR value per artery i.e. they don't provide information about the stenosis location or treatment strategy.
To address these, we propose a deep learning-based method to predict the FFR along the artery from CCTA scans. 
This study includes CCTA images of 110 patients who underwent invasive FFR pullback measurement in 112 arteries.
First, a multi planar reconstruction (MPR) of the artery is fed to a variational autoencoder to characterize the artery, i.e. through the lumen area and unsupervised artery encodings. Thereafter, using these artery characteristics as input, a convolutional neural network (CNN) predicts the FFR along the artery. The CNN is supervised by multiple loss functions, notably a loss function inspired by the Earth Mover's Distance (EMD) to predict the correct location of FFR drops  and a histogram-based loss to explicitly supervise the slope of the FFR curve.
To train and evaluate our model, eight-fold cross-validation was performed. The resulting FFR curves show good agreement with the reference allowing the distinction between diffuse and focal CAD distributions in the majority of cases. Quantitative evaluation yielded a mean absolute difference in the area under the pullback curve (AUPC) of 1.7 between the predicted FFR curves and the reference. The method may pave the way towards fast, accurate, fully automatic prediction of FFR along the artery from CCTA and may additionally differentiate diffuse and focal disease.
\end{abstract}


\section{Introduction}

Coronary artery disease (CAD) poses a significant threat to cardiac health \cite{roth_global_2017,roth_global_2018}. 
It is caused by the buildup of plaque in the coronary arteries, which can lead to stenosis, i.e. a luminal narrowing.
To determine the presence of stenosis, coronary CT angiography (CCTA) is typically visually assessed.
While visual assessment allows accurate evaluation of the anatomical grade of stenosis, it is associated with low specificity in the evaluation of functional stenosis, i.e. stenosis that hampers blood supply to the myocardium.
However, this is important, as only functionally significant stenoses require invasive revascularization \cite{pijls_fractional_2010,pijls_functional_2013}.
To determine the functional significance of a stenosis, fractional flow reserve (FFR) measurement is used as reference standard. In current clinical guidelines, FFR is measured invasively in the catheterization laboratory \cite{neumann_2018_2019,fihn_2014_2014}, which is costly and burdensome for patients.
In addition, due to the low specificity of the current CCTA-based patient selection, about 50\% of patients undergo invasive measurements although they do not require invasive treatment \cite{ko_combined_2012}.
To improve patient selection and prevent unnecessary invasive measurements, non-invasive FFR analysis from CCTA is an area of intensive research.

Several algorithms have been proposed for non-invasive prediction of FFR from CCTA. Currently, the most successful methods are based on computational fluid dynamics (CFD) \cite{taylor_computational_2013,norgaard_diagnostic_2014,tesche_coronary_2017,tesche_coronary_2018,coenen_diagnostic_2018,von_knebel_doeberitz_impact_2019,driessen_comparison_2019}. CFD-based methods derive the pressure along the coronary artery. 
However, these methods are computationally expensive. In addition, they require an accurate segmentation of the artery lumen and accurate determination of boundary conditions, which remains challenging. Occasional errors in the automatically obtained artery segmentation often require substantial manual interaction, and thus further extension of the total analysis time \cite{peper_diagnostic_2022}.

Alternatively, machine learning approaches have emerged for prediction of FFR from CCTA. 
Some of these methods make use of quantitative indices, e.g. stenosis degree and plaque volume, derived from CCTA \cite{ko_asla_2015,itu_machine-learning_2016,dey_integrated_2018,coenen_diagnostic_2018,von_knebel_doeberitz_coronary_2019,otaki_value_2021,yang_ct_2021}. These indices are typically fed to traditional machine learning algorithms to predict the presence of a functionally significant stenosis.
Although the indices correlate well with the measured FFR value, the ability of such models to capture complex relationships between FFR and coronary artery characteristics in CCTA may be limited.
Moreover, these methods, like CFD approaches, use semi-automatic coronary artery lumen segmentation, which can be a challenging and time-consuming task.


To address these limitations, deep learning approaches have been proposed as a fast yet analytically flexible solution for FFR prediction. Unlike conventional machine learning algorithms that require manual feature engineering, deep learning approaches automatically learn the features from the data. This potentially enables circumventing laborious correction of artery lumen segmentation.
Previous deep learning models typically first localize the arteries \cite{zreik_deep_2018,zreik_combined_2021,hampe_deep_2022} and extract features associated with functionally significant stenosis \cite{zreik_deep_2018,kumamaru_diagnostic_2020,zreik_combined_2021,hampe_deep_2022}, to enable training with the limited amount of available data. Finally, conventional machine learning algorithms \cite{zreik_deep_2018}, linear layers \cite{zreik_combined_2021} or a convolutional neural network (CNN) \cite{kumamaru_diagnostic_2020,hampe_deep_2022} to predict the presence of a functionally significant stenosis.
While performance of deep learning methods grows increasingly competitive with CFD methods, unlike the CFD methods, machine learning and deep learning models provide only a single FFR value per patient \cite{zreik_deep_2018,kumamaru_diagnostic_2020,zreik_combined_2021} or per artery \cite{zreik_deep_2020,hampe_deep_2022}. 
This value gives information about the total artery pressure drop, and hence about the necessity of treatment. However, it does not provide information about the spatial distribution of the drops, i.e. the location of the stenosis that requires the treatment.
However, clinically used invasive FFR can determine the FFR values along the artery \cite{collet_measurement_2019}.
This information is beneficial to locate the lesion that requires treatment. Furthermore, FFR measurements along the coronary artery may help in determining the treatment strategy.
Namely, stent placement is beneficial in case of a lesion with a focal drop in FFR values, while it is not suitable in case of diffuse disease where the FFR values drop gradually along the artery \cite{williams_patients_2010}.

Therefore, we propose a deep learning-based method for fast non-invasive assessment of the FFR along a coronary artery from CCTA.
First, we extract the coronary artery tree \cite{hampe_graph_2021}.
Thereafter, we build on our previous method where we designed a deep learning method that predicted a single FFR value per artery from CCTA \cite{hampe_deep_2022}.
The method consists of two stages, where in the first stage, we reduce the dimensionality of the input image by characterizing the artery along its length using supervised learning. In the second stage, we employ another deep learning network to predict the FFR value.
To enable stable training despite the relatively limited amount of training data, we maintain the two-stage architecture in this work. However, as prediction of FFR values along the artery is arguably a much harder task than predicting a single value, we improve the artery characterizations by combining supervised and unsupervised learning for extracting artery characteristics. This enables extraction of non-hand-crafted features from the artery, while explicitly incorporating anatomical information into the model through supervision with the lumen and plaque segmentation during training.
In the second stage, we use the combined characteristics as input to another deep learning network that first predicts the drop in FFR along the artery. This design is inspired by the natural process in which flow resistance accumulates along the coronary artery.
To train the second network for prediction of all FFR values along the artery, we make use of invasive FFR pullback measurements.
To supervise the prediction of FFR drops along the artery with the FFR pullback reference, we propose a novel loss function inspired by the Earth Mover's Distance (EMD) \cite{rubner_earth_2000}. Our EMD loss compares entire distributions of measurement points, as opposed to treating them separately, which implicitly accounts for uncertainty in the location of the reference drops. Additionally, to enable distinguishing between focal and diffuse CAD, we explicitly supervise the shape of the predicted FFR drops using a histogram-based loss.
Finally, we accumulate the FFR drops to obtain the FFR along the artery.



To the best of our knowledge, this is the first deep learning-based method for automatic prediction of FFR along the artery from CCTA allowing computation on a standard workstation without the need for cumbersome manual correction of the artery lumen segmentation during inference. 



\section{Data}
In this section we introduce the data that we used in this study: CCTA scans, invasively measured FFR values and semi-automatically obtained artery segmentations used for intermediate supervision during training.

\subsection{Patients and CT data}
Retrospectively collected CCTA scans of 118 patients (age 47-79 years) that underwent pullback FFR were included.
Scans were acquired at the Onze Lieve Vrouwe Ziekenhuis, Aalst, Belgium with a Siemens Somatom Definition Flash CT scanner.
The tube voltages ranged between 70 and 140 kVp and tube currents between 71 and 901 mAs. All scans were reconstructed to an in-plane resolution ranging from 0.28 to 0.70 mm$^2$ with 0.25 to 1.0 mm slice increment and 0.5 to 1.0 mm slice thickness.
During acquisition, contrast medium was injected with a flow rate of 4 to 6 mL/s for a total of 30 to 77 mL iopromide (Ultravist 300 mg I/mL, Bayer Healthcare, Berlin, Germany), depending on the patient weight and test bolus images. None of the patients underwent stenting or coronary artery bypass grafting prior to CCTA acquisition.
This study was approved by the institutional review board.
The inclusion criterion was that all arteries were in the field of view of the CCTA scan. 
Only scans of which the field of view contained all coronary arteries were included. 
In total, 8 of the 118 patients were excluded due to inadequate scan quality, e.g., severe step-and-shoot artifacts (n=6), artifacts caused by metal implants (n=1) or severe noise (n=1). This yields a total of 110 patients for further analysis.

\subsection{FFR pullback measurements}
All patients underwent invasive FFR measurements performed in at least one artery, which yielded in total 112 measurements. 
Measurements were made using a Certus Pressure Wire from St. Jude Medical. 
Adenosine was given through a vein to increase blood flow, and FFR was measured using a pressure guidewire in the coronary vessel. A motorized pullback was performed to obtain the FFR along the artery, with the device moving at 1 mm/sec until it reached the tip of the guiding catheter. The maximum time between a CCTA scan and the FFR measurement was 90 days.

\subsection{Reference artery segmentation}
\label{sec:refChar}
To facilitate training of artery characteristics extraction, artery segmentations are needed. For this, reference segmentations of the coronary artery lumen and calcified and non-calcified plaque were defined by manual annotation.

Coronary artery lumen segmentation in CCTA scans is a highly challenging task due to the small scale of the structures, the variability in contrast enhancement in the lumen, and blooming caused by the contrast agent. On the contrary, in X-ray angiography (XRA) images the coronary lumen is better visible as the resolution is higher and the contrast agent is easily distinguishable from plaque. Therefore, to manually annotate the coronary artery lumen in CCTA, corresponding XRA images were inspected. To this end, a 3D XRA model of the artery lumen was reconstructed from two XRAs acquired with different projection angles \cite{masdjedi_validation_2019}. XRAs were available for 72 arteries in 72 randomly selected patients.
 
To reduce the manual segmentation workload, an initial segmentation in CCTA was automatically generated \cite{wolterink_graph_2019}. Thereafter, the segmentation was transferred to the multi planar reconstruction (MPR) of the artery, which has 0.1 mm in-plane resolution and 0.5 mm slice distance. Subsequently, the segmentation was manually corrected in the MPR while simultaneously inspecting the artery lumen in 3D XRA.


Using the manual segmentations, the reference lumen area was computed by summing up the pixels of the respective segmentation in each cross-sectional slice of the MPR. Note that the MPRs for all arteries share the same spacings and in-plane resolution. Therefore, the area can be straightforwardly obtained by counting pixels.



\section{Methods}

\begin{figure*}[h!]
\begin{center}
\includegraphics[width=\textwidth]{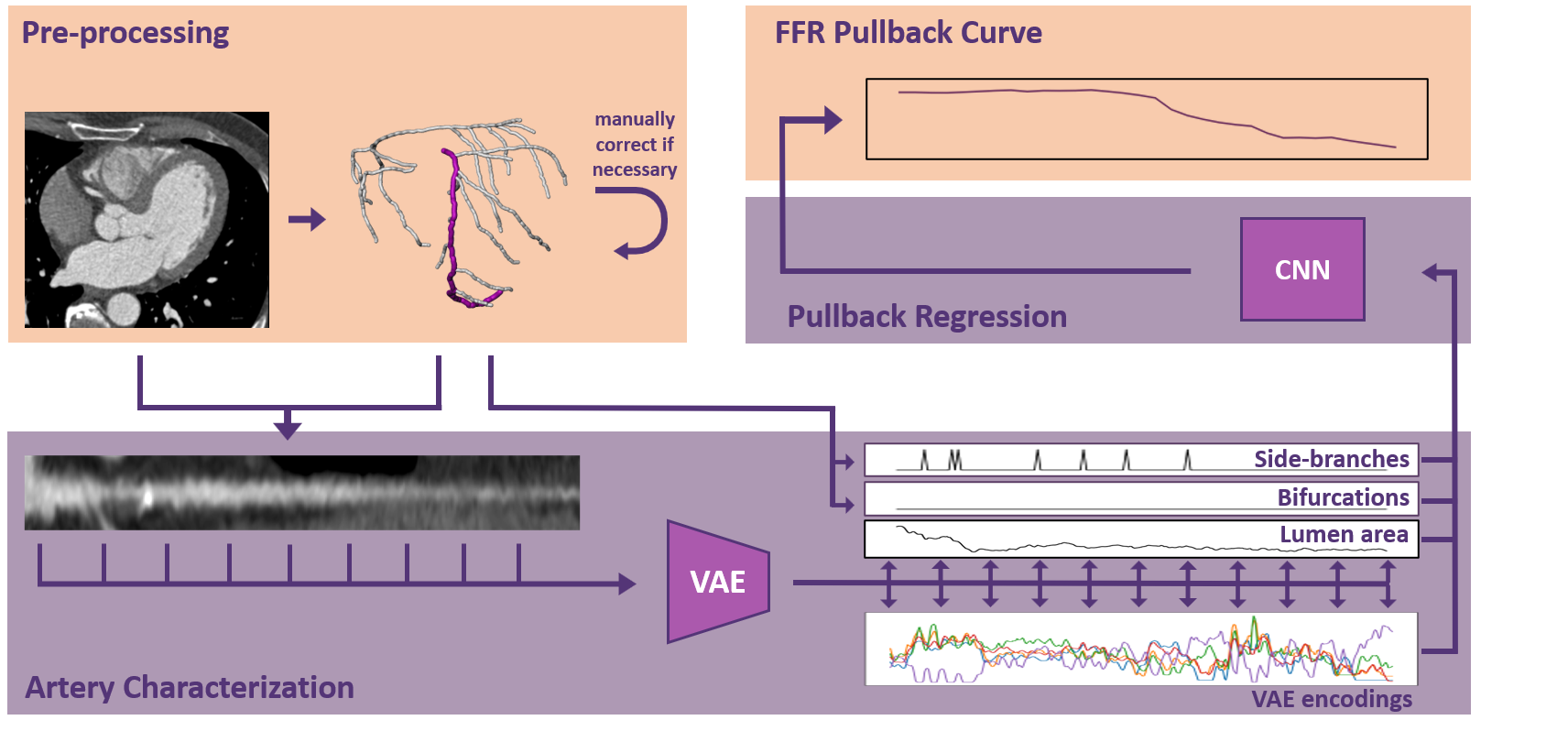}
\end{center}
\caption{Overview over our method. We first extract the centerlines of the coronary artery tree from the CCTA and manually correct them if necessary.
To predict FFR values along the artery, we analyze its MPR. This image is fed to the artery characterization network to regress the lumen area and generate VAE encodings. Subsequently, we merge these characteristics with characteristics directly extracted from the coronary artery tree. We use the merged characteristics as input to the pullback regression network, which predicts the FFR along the artery.}
\label{fig:methOverv}
\end{figure*}

Our method predicts the FFR values along a coronary artery from CCTA. 
To achieve this, first we extract the coronary artery tree. Thereafter, we compute characteristics from the coronary tree and characteristics from the artery and its immediate vicinity. To eliminate irrelevant information, we reconstruct an MPR for the arteries of interest from the tree. 
Thereafter, we use a variational autoencoder (VAE) \cite{kingma_auto-encoding_2013} to characterize the artery per cross-section of the MPR. 
To include information about bifurcations and side-branches, we extract additional characteristics directly from the coronary centerline tree. 
Finally, we predict the FFR values along the artery from the extracted characteristics using a 1D neural network (Figure \ref{fig:methOverv}).

\subsection{Coronary artery tree extraction} 

We use our previously developed automatic method to extract and label the coronary artery centerline tree from a CCTA scan using CNN-based tracking initialized at automatically placed seed points, and subsequent tree labeling based on graph convolutional neural networks. Subsequently, we inspect and manually correct the tree if necessary \cite{hampe_graph_2021}.
For each selected centerline, we then reconstruct MPRs with 0.1 mm in-plane resolution and 0.5 mm slice distance using trilinear interpolation, and normalize image intensities to maintain stability during neural network training.

\subsection{Artery characterization}

\begin{figure*}[h!]
\begin{center}
\includegraphics[width=\textwidth]{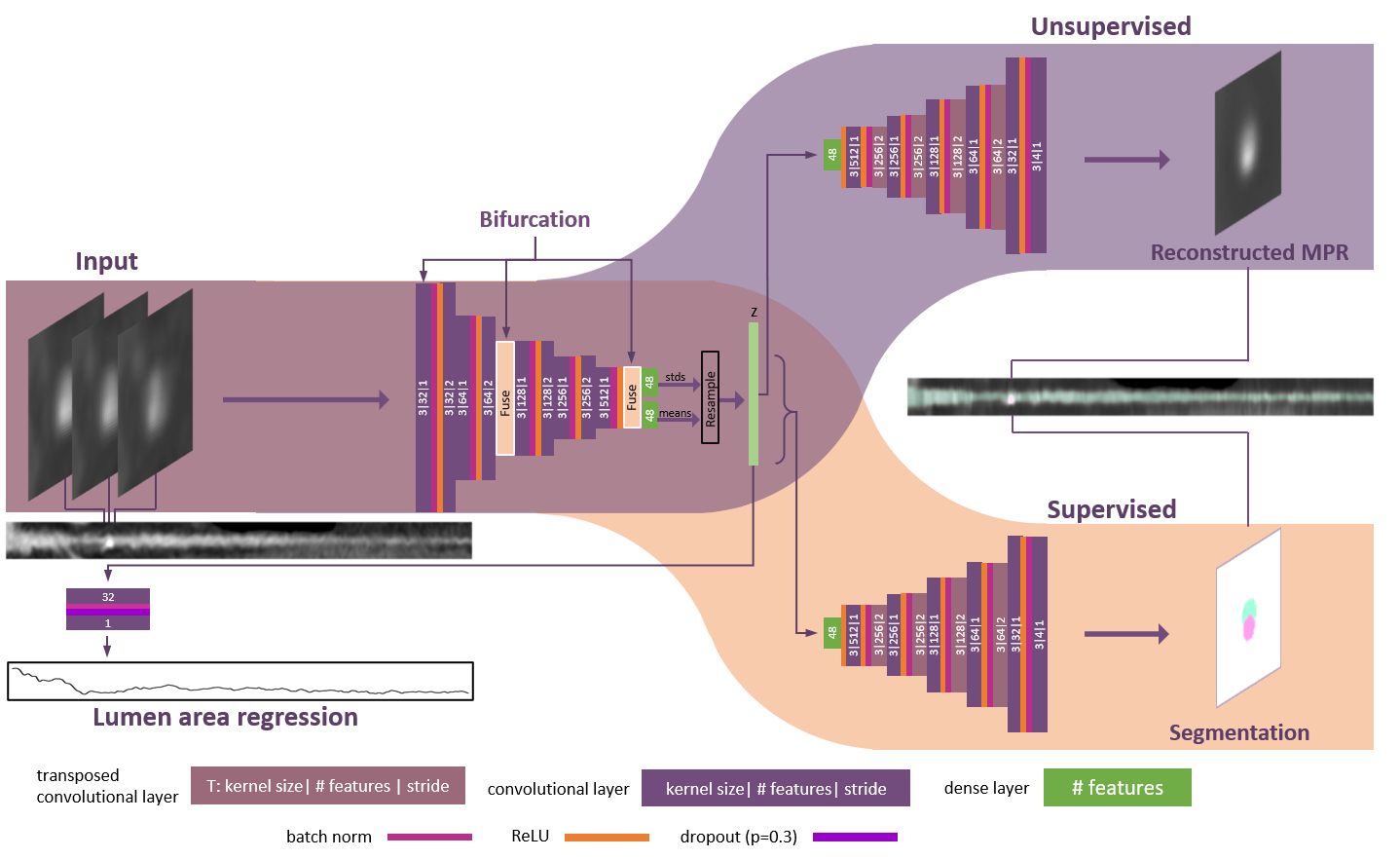}
\end{center}
\caption{VAE used for characterizing the coronary arteries. Stacks of 3 cross-sections are encoded using a convolutional encoder. The encoder consists of 5 convolutional blocks, each comprising 2 convolutions. To incorporate additional information about bifurcations, before the first, the third, and after the last convolutional block, a binary feature map is concatenated that indicates whether the MPR cross-section is located at a bifurcation. In the bottleneck, the latent vector $z$ is sampled from the predicted feature vectors $\mu$ (mean) and $\sigma$ (standard deviation). Subsequently, the first decoder reconstructs the central image of the input stack to ensure that the contained information is included in the encodings. To enhance the encodings with anatomical information, an auxiliary decoder receives only a subset of the features as input and reconstructs the segmentations for the lumen and the calcified and non-calcified plaque. Additionally, the features in $z$ are used as input to a linear layer that directly regresses the lumen area characteristic.}
\label{fig:methChar}
\end{figure*}

To characterize the coronary artery, we use a VAE, illustrated in Figure \ref{fig:methChar}. A VAE enforces latent features with independent normal distributions, increasing the interpretability and denseness of the latent space with respect to a conventional convolutional autoencoder (CAE) \cite{kingma_auto-encoding_2013}. 
First, we use a convolutional encoder to calculate features $z$ from a stack of three cross-sectional slices of the MPR, thereby incorporating information from adjacent slices. 
The encoder consists of convolutional blocks, containing two convolutional layers, batch normalization, and the ReLU activation. The second convolutional layer in each block has a stride of 2, which enables downsampling the resolution. To inject additional context information, after the second convolutional block the output of the network is concatenated ("fused") with the feature maps resulting from applying the preceding layers to the 8 MPR cross-sections adjacent to the input cross-sections. After the fourth convolutional block, the same is done using the 2 adjacent MPR cross-sections. This increases the receptive field in the z direction to 13 cross-sections or 6.5 mm. Furthermore, outgoing arteries at bifurcations can appear as luminal widening. To reduce potentially associated lumen overestimation, information about bifurcations, obtained from the coronary artery centerline tree, is injected before the first, the third, and after the last convolutional block. For this, a feature map indicating whether the MPR cross-section is located at a bifurcation is concatenated with the features at these locations. 

To extract relevant information from the input image into the features, a decoder is trained to reconstruct the central image of the input stack from the feature vector $z$ (top pathway in Figure \ref{fig:methChar}). Like the encoder, the decoder consists of alternating convolutional blocks and pooling operations. However, for the decoder, the last convolutional layer in each block is transposed to enable up-sampling. Furthermore, only a single image is reconstructed instead of the whole input stack.

To enhance a subset of the features with information about artery and plaque geometry, an auxiliary output decoder predicts segmentation masks for the lumen, the calcified and the non-calcified plaque (bottom pathway in Figure \ref{fig:methChar}). The architecture of the segmentation decoder is equivalent to that of the VAE decoder, except for the last layer, which has four channels and uses the softmax activation.

To surpass the decoders at testing, thereby circumventing the challenging artery segmentation and enabling fast computation, we directly regress the artery lumen area from the encoded features. This artery lumen characteristic is utilized in further analysis. Moreover, the 32 predicted means corresponding to the subset of features used in the segmentation decoder (bottom path) are also used in further analysis. 
The remaining 12 encodings do not contain information used by the segmentation decoder and hence, likely describe the background. Therefore, we discard these encodings. 

While a reduction in lumen area can indicate the start of a lesion, it can also be caused by outgoing bifurcations and transitions from the main into a side-branch.
To explicitly distinguish these causes we extract two additional, binary characteristics at the level of a given centerline point by directly using the coronary centerline tree. 
One characteristic indicates the locations of bifurcations with a 1, whereas the values at other centerline points remain 0. Another characteristic marks centerline points in side-branches with a 1, while points in the main branches (e.g. left main, LAD, LCX, RCA) remain 0.

To ensure consistency across our training data set, we normalize all characteristics to zero mean and unit variance. Thereafter, binary characteristics are capped at 1.

\subsection{Prediction of FFR values along the artery}

To predict the FFR from the extracted characteristics we first lay out the architecture of the FFR prediction network.
To enable accurate prediction of the FFR along the artery, we supervise this network with the reference FFR drop. 
However, this is challenging as correct registration of the reference FFR pullback with the input artery can not be guaranteed. To nevertheless predict the correctly shaped FFR drops at the right locations, we combine two separate loss functions, which are described in the subsequent sections. 


\subsubsection{FFR prediction network}
\label{sec:meth:FFRPred}

\begin{figure*}[h!]
\begin{center}
\includegraphics[width=\textwidth]{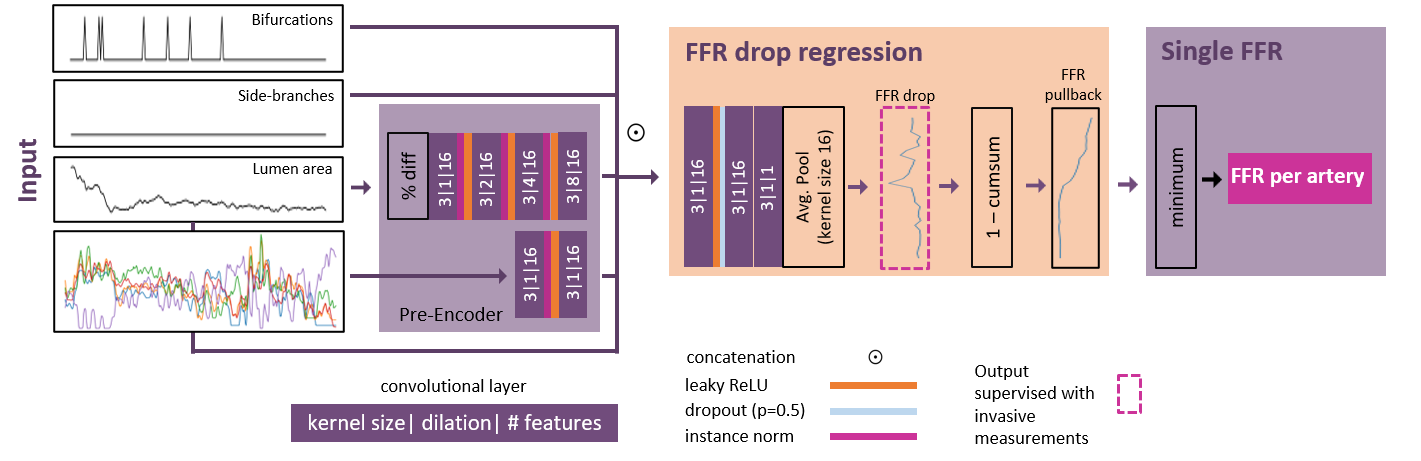}
\end{center}
\caption{
Network architecture for prediction of FFR along the artery. The network consists of an encoder and one output head for regressing the FFR per centerline point. We first pre-encode the lumen area and the VAE encodings. We combine the results with the remaining characteristics and feed them to the regression head. To calculate the FFR drop along the artery, the regression head comprises two convolutional layers, average pooling, and the ReLU activation function. Subsequently, we calculate the FFR along the artery by taking the cumulative sum over the predicted FFR drops and subtracting each value from one. Lastly, we extract the FFR value at the artery level by calculating the minimum over the predicted FFR values. 
}
\label{fig:methSten}
\end{figure*}

We use a 1D CNN to analyze the extracted artery characteristics and assess the FFR (Figure \ref{fig:methSten}). 

Before analyzing the combination of the extracted characteristics, we apply convolutions to them individually. Thereby, for the lumen area, which is typically larger proximally than distally, a challenge arises, as convolutions are applied to all locations equally, i.e. assuming translation equivariance of features. Therefore, to minimize the impact of the descending lumen area, we calculate the percentage difference of each value in the artery with the previous one. For this we, divide each value by the preceding value and subtract the result from 1, the FFR value at the start of the artery. Subsequently, we pre-encode the lumen area characteristic and the VAE encodings separately to encode larger structures like lesions.
Specifically, for the lumen area characteristic, we apply four convolutional layers with rising levels of dilation to enlarge the receptive field to encompass entire lesions. For the VAE encodings, we use only two convolutional layers, to prevent potential overfitting. For the same reason, we only include the VAE features that were attached to the segmentation decoder, as these likely describe the artery instead of the here irrelevant background. 
Thereafter, we introduce a skip connection by concatenating the resulting features with the input characteristics and 
feed the resulting encodings to a common convolutional pathway for regression of the FFR along the artery. 
Inspired by the additive nature of sequential flow resistances, the regression pathway first predicts the FFR drop per point in the artery. In this way, the prediction target at each point is independent of the previous outputs. This enables their prediction using local features, as computed by the convolutions. 

The network for prediction of FFR along the artery uses a kernel size of 3 for all convolutions with zero-padding to preserve feature size. Furthermore, we utilize 16 filter maps to balance expressiveness and overfitting. To further prevent overfitting, we set dropout probabilities to 0.5 and instance normalization is used throughout the network. 

To mitigate the impact of potential misregistration between the input artery and the reference FFR drop per point on training, we make use of average pooling. Average pooling combines adjacent outputs, rendering small location differences irrelevant. 
As the FFR is typically measured 10 mm distally to the annotated lesion location, but the extracted MPRs may be longer, we mask all predicted FFR drops distal to the typical measurement location.
To obtain the FFR values along the artery from the predicted FFR drops, we calculate the cumulative sum over the FFR drops.

The here predicted FFR values along the artery can be used to calculate a single FFR value for the entire artery, which directly indicates the need for invasive treatment. Analogously to the clinical workflow, this single FFR value is obtained by computing the minimum of the FFR values along the artery.


\subsubsection{Positional loss}
\label{sec:EMD}
\begin{figure*}[h!]
\begin{center}
\includegraphics[width=0.9\textwidth]{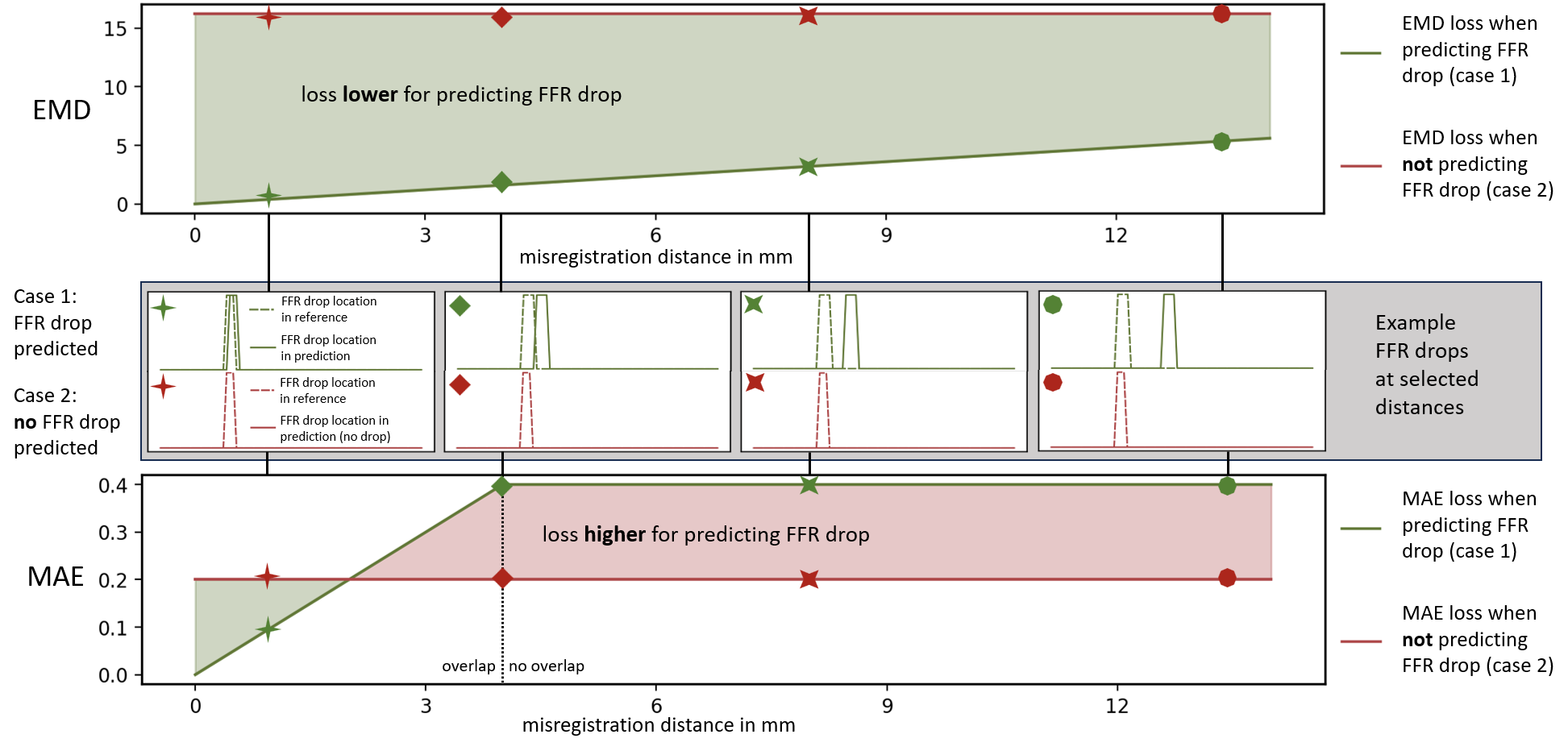}
\end{center}
\caption{
Behavior of the proposed EMD loss compared to the commonly used MAE loss, demonstrated with a varying level of misregistration for different prediction situations. In this figure synthetic data is used to simulate the behavior of the loss functions. We distinguish two possible cases in presence of a reference drop: predicting an FFR drop in the correct location (green) and erroneously predicting no FFR drop (red). Misregistration can cause a lower MAE for erroneously predicting no drop (case 1) than predicting an FFR drop at the misregistered yet correct location (area shaded in red). In contrast, the EMD increases continually with the distance between the drops but crucially, remains lower than the EMD associated with predicting no FFR drop. Please note that we equalize the different loss magnitudes with the loss weightings in our experiments.
}
\label{fig:EMD}
\end{figure*}

To predict the correct location of the FFR drops, we propose a loss function inspired by the EMD \cite{rubner_earth_2000}. Instead of locally comparing two probability distributions, EMD is a way to measure the global similarity between the distributions.
Intuitively, EMD is the minimum amount of "work" required to transform one distribution into another. Thereby, "work" is defined as the amount of probability mass that needs to be moved, multiplied by the distance it needs to be moved. 
Therefore, unlike a point-wise comparison like the mean absolute error (MAE), this loss function increases continuously with the distance between the predicted FFR drop and the reference FFR drop. Whereas the MAE remains the same for increasing distances between the reference and the predicted FFR drop, the EMD increases. 
This potentially yields improved gradient updates in case of misregistration between the predicted and the reference pullback (Figure \ref{fig:EMD}). 
In a single dimension, the EMD between two probability distributions $p_1$ and $p_2$, discretized via $x_i$ for $i \in [0, l]$, is calculated as

\begin{equation}
    \textrm{EMD} = \sum_{j=0}^{l} \left| \sum_{i=0}^{j} \left( p_1(x_i) - p_2(x_i)\right) \right|
\end{equation}

Although originally defined for probabilities, in this work we use the above formula to calculate the loss between the predicted and the reference FFR drop. Intuitively, this calculation corresponds to the accumulated difference between the reference FFR pullback and the predicted FFR curve:

\begin{align}
    L_{\textrm{FFR}} &= \sum_{j=0}^{l} \left| \sum_{i=0}^{j} \Delta FFR_{pred}(x_i) - \Delta FFR_{ref}(x_i) \right| \\
   &= \sum_{j=0}^{l} \left| FFR_{pred}(x_j) - FFR_{ref}(x_j) \right| \quad .
\end{align}



However, since the calculation of the FFR from the FFR drop is asymmetric, $L_{\textrm{FFR}}$ punishes the differences in the proximal FFR drop more than the distal ones. Instead, to treat differences in FFR drops equally regardless of their location, we design a symmetric version of $L_{\textrm{FFR}}$,  $L_{\textrm{FFR}}^{symm}$. For this, we add a second term $\bar{L}_{\textrm{FFR}}$ calculated from a hypothetical FFR curve computed by adding up FFR drops from distal to proximal:

\begin{align}
    L_{\textrm{FFR}}^{symm} &= L_{\textrm{FFR}} + \bar{L}_{\textrm{FFR}} \quad where\\
    \bar{L}_{\textrm{FFR}} &= \sum_{j=0}^{l} \left| \sum_{i=0}^{j} \Delta FFR_{pred}(x_{l-i}) - \Delta FFR_{ref}(x_{l-i}) \right| \quad .
\end{align}

\subsubsection{Histogram loss}
\begin{figure*}[h!]
\begin{center}
\includegraphics[width=0.75\textwidth]{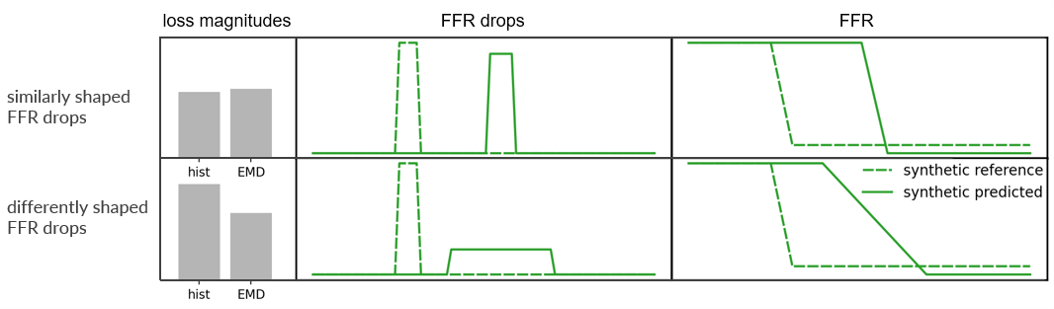}
\end{center}
\caption{Utility of the histogram loss for supervising the shape of FFR drops, compared to the EMD loss. 
The Earth Mover's Distance (EMD) loss and the histogram loss serve distinct roles in our method. 
On the left side, the loss magnitudes are shown that were calculated from the FFR drops in the middle column. On the right side, the corresponding FFR curve is depicted.
While the EMD loss is sensitive to the overall mass displacement between the predicted and reference FFR drops, in case of no overlap between the drops it does not penalize variations in shape. In contrast, the histogram loss is introduced specifically to address this, penalizing differences in shape between the predicted and the reference FFR drops. 
However, the histogram loss is invariant to the location of the FFR drops. Therefore, a combination of both the EMD and the histogram loss is needed to ensure the correct shape of FFR drops at their accurate locations.
}
\label{fig:hist}
\end{figure*}

To enable distinguishing focal from diffuse CAD, it is crucial that the shape of the predicted FFR drops is similar to that of the FFR drops in the reference FFR pullback (Figure \ref{fig:hist}). 
Therefore, we explicitly supervise the shape of FFR drops by penalizing the differences between the histogram of the predicted FFR drops and the reference. However, straightforwardly binning output values is not differentiable and therefore does not enable training of a neural network. To solve this problem, we instead use a Parzen window approach by approximating the bins using normal distributions with means located at the respective bin's center. 
We then penalize the differences between these approximated histograms using the weighted absolute error. 
We weight each bin according to the magnitude of the corresponding FFR drop, giving more weight to larger FFR drops, which are less common but clinically more important.
Figure \ref{fig:hist} demonstrates, that the proposed histogram loss increases for increasing shape differences between the predicted and the reference FFR drops. Please note, that as the histogram loss is invariant to the location of FFR drops, the combination of EMD and histogram loss is required to obtain the correctly shaped FFR drops at the right locations.

\section{Evaluation}


To evaluate the FFR along the artery, we compare the predicted FFR values at each point in the artery with the reference values. For this, we compare the distributions using histograms and the individual FFR values using scatter plots. Furthermore, we calculate the area under the pullback curve (AUPC) by first interpolating the predicted and the reference pullback to a spacing of 10 mm \cite{collet_impact_2018}. Subsequently, we obtain the AUPC by adding up all FFR values.

To assess the shape of the predicted FFR along the artery, we calculate the pullback pressure gradient (PPG) index \cite{collet_measurement_2019}. The PPG index is calculated as 

\begin{align}
    \textrm{PPG index} = \frac{\left\{ \frac{MaxPPG_{20mm}}{\Delta FFR_{vessel}} + \left( 1 - \frac{\textrm{Length with functional disease (mm)}}{\textrm{Total vessel length (mm)}}\right) \right\}}{2}
\end{align}

The first term describes the maximum gradient over a window of 20 mm, divided by the total FFR drop over the artery (between the FFR value at the ostium and the minimal FFR value). The second term quantifies the proportion of the artery with functional disease. Here functional disease is defined by an FFR drop $\geq 0.0015/mm$.
A high value for the PPG index indicates that FFR drops are concentrated into small regions (focal CAD). 

In addition to assessing the PPG value of an artery, we classify the artery as focal vs. diffuse disease using the PPG. Specifically, analogously to Sonck et al. \cite{sonck_clinical_2022}, we use the median of all reference PPG values in our dataset (PPG$_{thres}$ = 0.63) as a threshold. Arteries with a PPG value above this threshold are classified as having focal and the remaining arteries as having diffuse disease. To assess the ability of our model to classify focal vs. diffuse disease, we calculate the accuracy, the sensitivity and the specificity by comparing the predicted with the reference classifications.

\section{Results}

To develop and evaluate our method, we applied eight-fold cross validation for FFR prediction. In addition to evaluating the proposed method, we performed ablation experiments to investigate the contributions of individual components.

\subsection{Experimental settings}
\begin{figure}[h!]
\begin{center}
\includegraphics[width=0.5\linewidth]{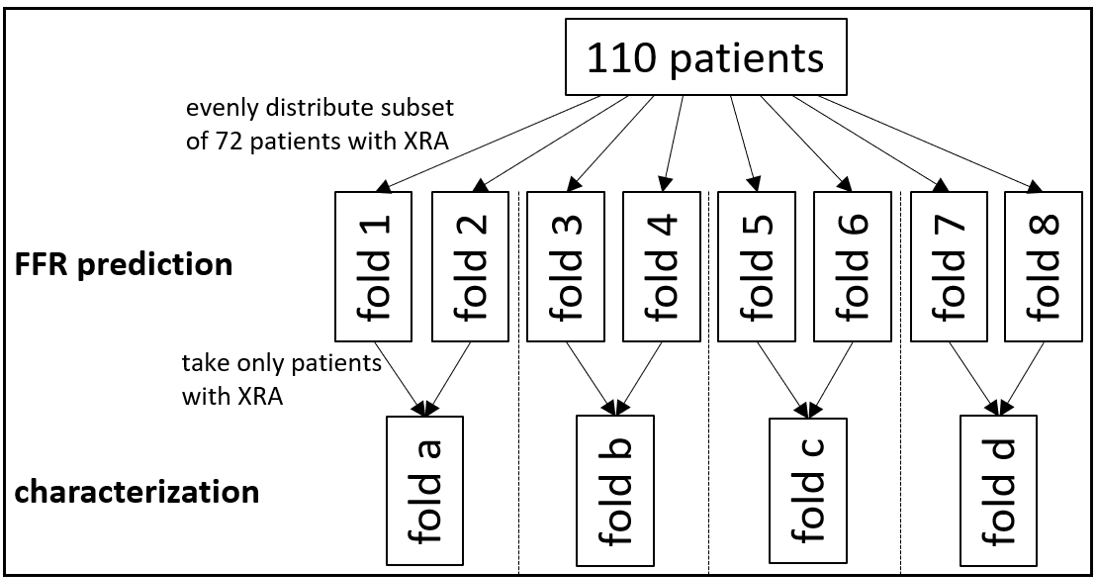}
\end{center}
\caption{Distribution of data across the folds for the FFR prediction network and the characterization network. To prevent leakage of training data into testing, we ensured that no characterizations of patients used in training of the characterization network were used in testing of the FFR prediction network. For example, the characterization network for which fold a was not used in testing was paired with the FFR prediction networks that were tested on folds 1 and 2.}
\label{fig:folds}
\end{figure}

To prevent leakage of test data into training, in testing we used each characterization network only on data that it was not trained on. For this, we distributed the training data for the characterization network across the eight folds used to train the FFR prediction network (Figure \ref{fig:folds}).
As training of the characterization network was computationally expensive, we trained four of these networks, each of which was used to yield input for two of the FFR prediction networks.
Each artery characterization network underwent 1,200 epochs of training. For supervision, we used the MAE as the reconstruction loss, binary cross-entropy as the segmentation loss, and the Kullback-Leibler divergence for regularization of the latent space.
To specifically improve the representation of the artery, we additionally weighted the reconstruction loss within the reference segmentation (lumen and plaque) with a factor of 5. To supervise the lumen area regression, the MAE was used, weighted with a factor of 10. The network was optimized using the ADAMW \cite{loshchilov_decoupled_2019} optimizer with a learning rate of $10^{-5}$ and a batch size of 512. Once trained, we applied the network to each cross-section of the MPR to extract the lumen area and the VAE encodings along the centerline. 

The network for FFR prediction was trained for 150 epochs, employing the ADAMW optimizer with a linearly scheduled cyclic learning rate. The cyclic learning rate varied between 5e-4 and 1e-5 over a period of 40 epochs. Due to the limitation imposed by different lengths of the arteries, the network could process only a single artery at a time. Therefore, the loss was accumulated over eight training iterations before backpropagating, corresponding to an effective batch size of 8.

To supervise the FFR along the artery, we counterbalanced fidelity and sensitivity to noise in the pullback reference by pooling the output of the network to a 2 mm step size using average pooling with a kernel size of 4. 
To minimize misregistration, we manually registered invasive pullback measurements with the input by shifting the beginning of the pullback so that FFR drops optimally overlap with lumen narrowings. This resulted in an average absolute shift of 4.5 mm.
As the pullback reference is only measured in the clinically relevant part of the artery and the MPR often continues more distally than the measurement location, we applied pullback supervision to the relevant part, by masking the distal remainder.
For the histogram loss, we employed 32 normal distributions with sigma 0.1, equidistantly distributed between -0.1 and 0.5, i.e. the expected range of FFR drops. 

Additionally, we regularized the network to incorporate the fact that the FFR typically decreases monotonously. Although negative FFR drops, corresponding to an increase in FFR, exist in the dataset, these are very rare. Preliminary experiments showed that preventing negative values for the FFR drops, i.e. by using a ReLU activation function as the last layer, leads to numerical instability in training. Therefore, we enable negative output but discourage it by penalizing the absolute of the sum of negative FFR drops.

The EMD loss, the histogram loss and the monotonicity regularization were weighted with factors 0.1, 5 and 20, respectively. The factors were chosen on the basis of preliminary experiments to achieve similar magnitudes for the loss terms.

\subsection{Prediction of FFR along the coronary artery}

\begin{figure*}[h!]
\begin{center}
\includegraphics[width=\textwidth]{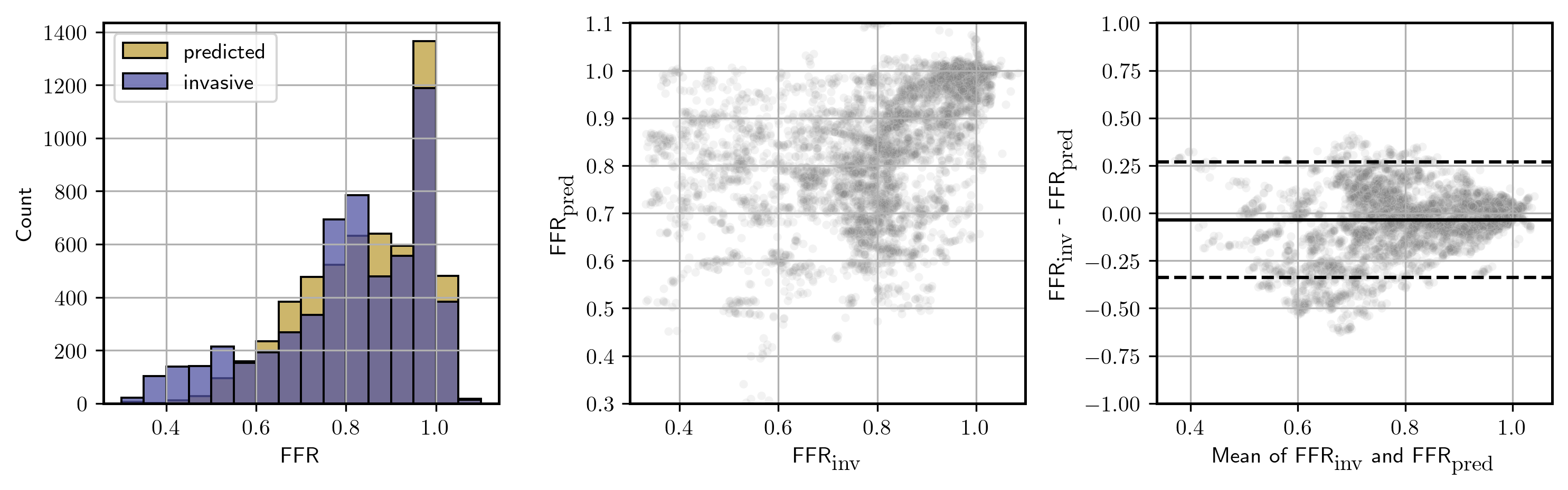}
\end{center}
\caption{Quantitative results for all points in the arteries (each point represents the FFR in one location in an artery). On the left, the histogram over the predicted FFR values is compared with the histogram over the reference FFR values. In the middle, a scatter plot allows direct comparison of all FFR values. On the right, a Bland-Altman plot is shown.}
\label{fig:perPoint}
\end{figure*}

First, the FFR prediction along the arteries is evaluated. The results are shown in Fig. \ref{fig:perPoint}, where each artery contributes as many data points as reference FFR measurements. On the left, histograms of the predicted and reference FFR values are shown. Results reveal 86 \% overlap between them indicating that the output of our method typically lies in the correct range. 
In the middle, a scatter plot directly compares predicted and reference FFR values at all available points of reference measurement.
As observed from the graph, predictions above approximately 0.7 tend to align well with the reference, whereas predictions below 0.7 tend to be overestimated by up to 0.6 in the worst cases. The threshold of 0.7 is not based on clinical considerations, but reflects an observation of the model's performance.
The observation is confirmed by the Bland-Altman plot on the right, which shows a trend to predicting too high FFR values only below 0.7.

\begin{figure*}[h!]
\begin{center}
\includegraphics[width=0.4\linewidth]{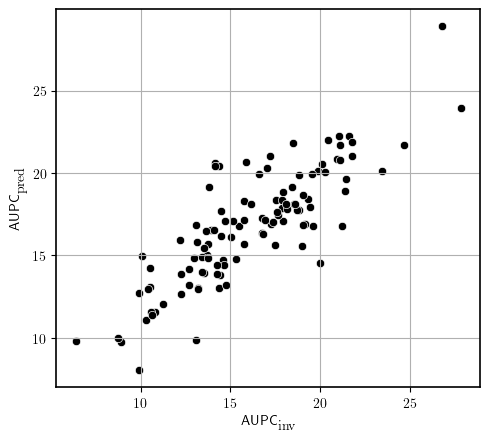}
\end{center}
\caption{Scatter plot comparing predicted and reference AUPC values. Each point corresponds to the area under one FFR curve.}
\label{fig:AUPC}
\end{figure*}

To analyze the predicted FFR curves per artery, the AUPC was calculated and compared with the reference in a scatter plot (Figure \ref{fig:AUPC}). The plot demonstrates good agreement between the predicted and the reference AUPC values. Quantitative evaluation yielded a mean absolute difference of the predicted and the reference AUPC values of 1.7.

\begin{figure*}[h!]
\begin{center}
\includegraphics[width=0.4\linewidth]{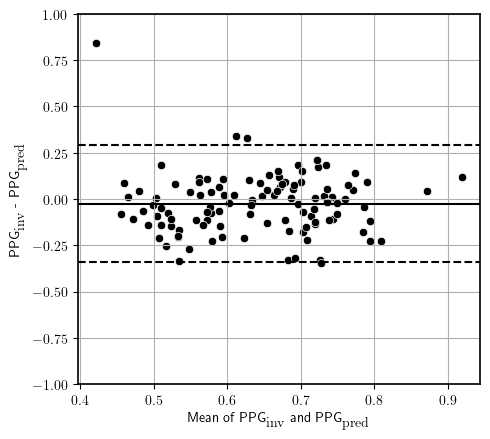}
\end{center}
\caption{Bland-Altman plot comparing the PPG index computed from the predicted FFR curve with the PPG index from the reference FFR pullback.}
\label{fig:PPG}
\end{figure*}

To assess focal vs. diffuse CAD, the PPG index was calculated for the predicted and reference FFR curve. 
Figure \ref{fig:PPG} shows a Bland-Altman plot of the resulting values demonstrating the absence of bias in the PPG index. 
To directly assess the ability of the model to distinguish focal from diffuse CAD, following Sonck et al. \cite{sonck_clinical_2022} we calculated the median PPG of the reference pullbacks as threshold to classify the disease in one of the two categories. In our dataset, this median equals 0.63. The resulting accuracy, sensitivity and specificity were 0.70, 0.80 and 0.60, respectively.

\begin{figure*}[h!]
\begin{center}
\includegraphics[width=0.8\linewidth]{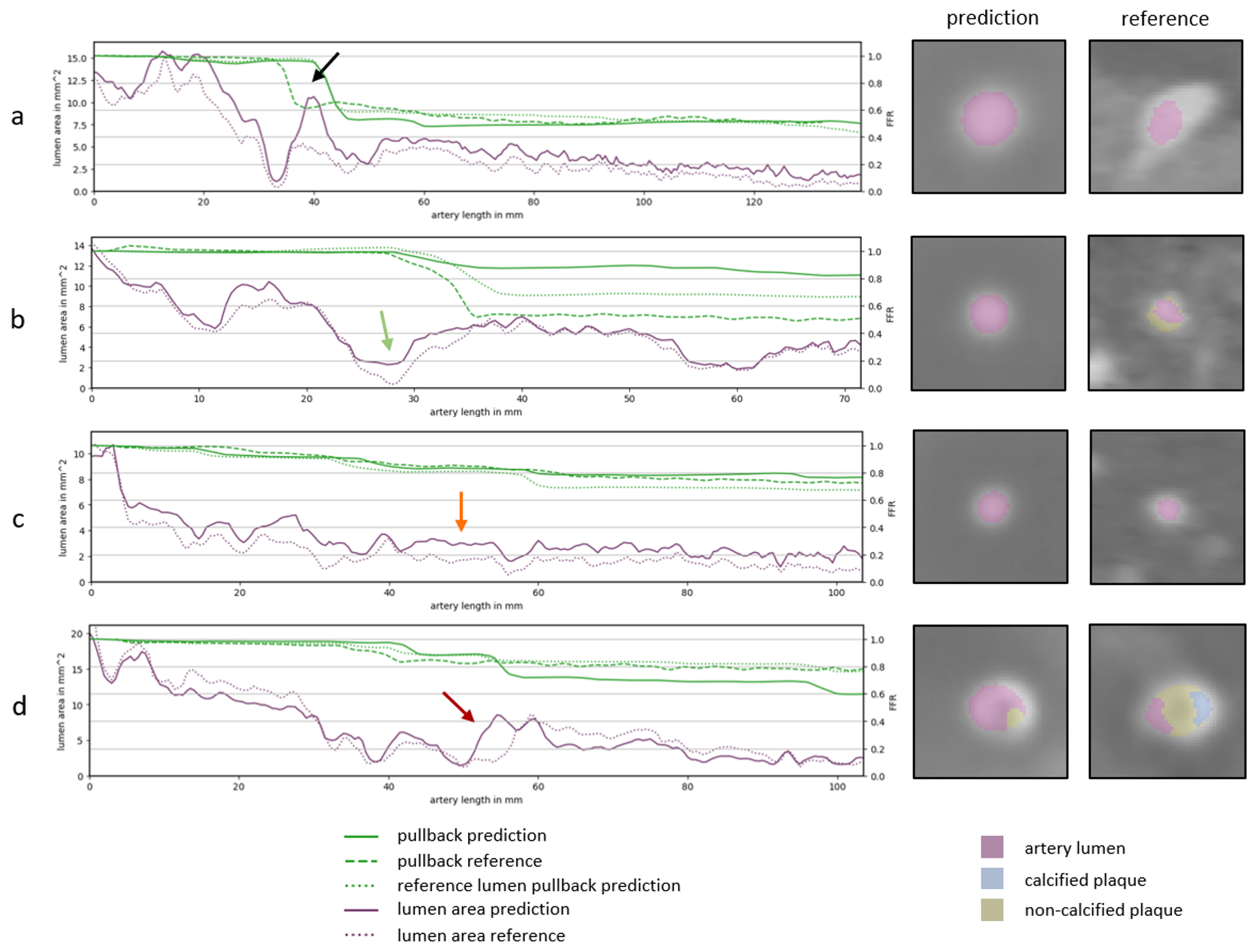}
\end{center}
\caption{Lumen (purple, left y-axis) and FFR curves (green, right y-axis) for 4 arteries. Predictions of our method are shown in solid lines and the reference FFR pullback is shown in a dashed line. The reference lumen is depicted in a finely dotted line, as well as the prediction that results using the reference lumen as input to our FFR prediction network. On the right, the output of the decoders, i.e. the reconstructed cross-sectional slice and the predicted segmentation, is compared to the reference segmentation overlaid on the input image. The position of the slice is indicated with an arrow in the graph. Arteries a and b correspond to focal CAD in the reference, and arteries c and d to diffuse CAD. Whereas in arteries a and c, the method correctly predicted focal and diffuse disease, respectively, in arteries b and d the predictions disagreed with the reference. In these arteries, the shape of the predicted lumen area is incorrect at the marked locations, whereas showing high agreement with the reference otherwise. Substituting the reference lumen area as input leads to the correct predictions of focal and diffuse disease in arteries b and d, respectively.}
\label{fig:curves}
\end{figure*}

In Figure \ref{fig:curves} predicted FFR curves in four representative examples are compared to the reference. To assess the relationship between the artery geometry and the FFR along the artery, the predicted and reference lumen area are shown as well. Furthermore, cross-sectional slices of the MPR with overlaid segmentations are shown on the right, to showcase typical errors (respective location in the graph indicated by an arrow).
Arteries a and b correspond to focal CAD, and arteries c and d to diffuse CAD.
The prediction of the lumen area is accurate throughout most of the arteries, although an offset is present in arteries a and c. 
In artery a, our method correctly identified the focal FFR drop. However, the lumen area is overestimated at the location indicated by the black arrow. As visible in the MPR cross-section, the reason for this is an outgoing bifurcation.
In contrast, in artery b the focal lesion was missed due to an underestimation of the luminal narrowing. 
The corresponding cross-sectional slice shows that the reason for this was failure to identify non-calcified plaque. 
Whereas artery c was identified by the method as having diffuse CAD, the prediction for artery d shows incorrectly detected focal components. Similarly to artery b, the error is caused by an overestimation of the lumen area, related to confusing non-calcified plaque for lumen.
To investigate whether improvement of the artery characteristics leads to a better FFR prediction performance in these cases, we substituted the reference lumen as input to the FFR prediction network. In arteries b and d this led to improved predictions.
Specifically, using the more severely dropping reference lumen area in artery b enabled predicting the associated focal FFR drop. In artery d, substituting the less steeply rising reference lumen area resolved the incorrectly predicted focal FFR drops.


\subsection{Ablation experiment}

To investigate the impact of our methodological choices, we performed ablation experiments to assess the performance of different algorithm configurations (Table \ref{tab:ablat}). Specifically, we examined the effects of using the MAE loss instead of the EMD loss, as well as of using each loss individually. 
While the proposed configuration utilizing the EMD loss aligns with the one using the MAE loss, it demonstrates a considerably greater sensitivity, implying that it predicts more focal drops. Applying each loss separately results in substantially lower accuracies.
Moreover, we evaluated the utility of the VAE features.
Eliminating the VAE features led to a loss in sensitivity, while specificity increased.
At last, we investigated the effect of including the VAE features that describe the background ("All VAE features"). The resulting accuracy and sensitivity are slightly below the values of the proposed approach, while the specificity remained the same.


\begin{table*}[htbp]
  \centering
  \caption{Results of ablation experiments in terms of accuracy, sensitivity, and specificity for diagnosing focal vs diffuse disease. 
  Although the proposed configuration using the EMD loss is matched by the configuration using the MAE loss, it shows a substantially higher sensitivity, meaning that more focal drops are predicted correctly when using the EMD loss vs when using the MAE loss. Using each loss individually yields substantially reduced performance in terms of accuracy. 
  Additionally, we evaluate the utility of the VAE features. Removing the VAE features leads to a higher specificity but also results in a substantial loss in sensitivity and accuracy. In the last experiment, instead of using only the 32 encodings for the VAE that were attached to the segmentation decoder, we used all 48 encodings. This leads to a similar performance as the proposed algorithm, however with slightly lower accuracy and sensitivity.}
  \resizebox{0.7\textwidth}{!}{%
    \begin{tabular}{lccc|rrr}
    \toprule
    \multicolumn{1}{c}{Configuration} & EMD & MAE & Hist & \multicolumn{1}{c}{Accuracy} & \multicolumn{1}{c}{Sensitivity} & \multicolumn{1}{c}{Specificity} \\
    \midrule
    Proposed & \checkmark & & \checkmark & \textbf{0.70} & \textbf{0.80} & 0.60 \\
    \arrayrulecolor{black!30}\midrule
    MAE instead of EMD & & \checkmark & \checkmark & \textbf{0.70} & 0.59 & 0.82 \\
    Only EMD & \checkmark & & & 0.56 & 0.27 & \textbf{0.85} \\
    Only MAE &  & \checkmark & & 0.62 & 0.61 & 0.64 \\
    Only Hist &  & & \checkmark & 0.55 & 0.34 & 0.76 \\
    \arrayrulecolor{black!30}\midrule
    No VAE Features & \checkmark & & \checkmark & 0.66 & 0.66 & 0.65 \\
    All VAE Features & \checkmark & & \checkmark & 0.68 & 0.77 & 0.60 \\
    \arrayrulecolor{black}\bottomrule
    \end{tabular}
  }
  \label{tab:ablat}
\end{table*}%

\section{Discussion}
Our study introduces a deep learning approach to predict the FFR along a coronary artery from CCTA scans. The method combines supervised and unsupervised characteristics extracted from the artery's MPR and from the coronary artery tree. 
Subsequently, using the extracted characteristics, the FFR drop at each point in the artery is regressed. By accumulating the FFR drops we obtain the final FFR along the artery.

The predicted FFR along the artery may be directly usable as visual guidance for the treating clinician, i.e. to derive the location of the culprit lesion or to visually assess whether a patient has focal or diffuse CAD. 
Additionally, we demonstrate the utility of the FFR curves for automatic classification of focal vs. diffuse CAD. For this, we calculated the PPG index and thresholded it with the median PPG index in the dataset.
The FFR curves predicted by our method allowed for an accurate differentiation between diffuse and focal CAD in most cases.

As opposed to other methods that use simulated data to obtain a large training set \cite{itu_machine-learning_2016,li_prediction_2021}, our deep learning network was trained using only invasive pullback measurements. Therefore, our model is more likely to generalize to real-world data. However, this comes with several challenges. 
First, the number of available invasive pullback measurements was limited. This is a challenge when training deep learning networks, as these are prone to overfitting. To address this, we used a two-step method where clinical knowledge is integrated into the first step. This was achieved by using the reference segmentations for supervision.
Second, because we combine two different modalities in training, i.e. CCTA images and invasively measured FFR pullbacks, misregistration can be present. To mitigate the impact of the possible misregistration on training, specific loss functions were designed that are robust to such errors. Despite these challenges, the results demonstrate that prediction of FFR along the artery from CCTA using deep learning is feasible without artery segmentation.

To highlight the impact of our methodological choices, multiple ablation experiments were performed.
We designed multiple loss functions to enable predicting focal FFR drops, even if the exact location of the drop is uncertain, i.e. due to potential uncertainties arising from factors like misregistration (Figures \ref{fig:EMD} and \ref{fig:hist}). The results show that, although enabling equivalent accuracy for distinguishing focal and diffuse drops, using the MAE instead of the EMD yields a significantly lower number of correctly predicted focal drops (lower sensitivity). This demonstrates the superiority of the EMD loss for the task at hand, as in the given clinical setting, false negatives are more problematic than false positives. For optimal performance the combination with the histogram loss is required.

Removing VAE features led to a drop in sensitivity and a slight increase in specificity.
This suggests that the VAE features contain valuable information for identifying obstructive lesions, albeit at the cost of some false positives. In addition to information about the artery lumen, the VAE features also include information about plaque, as plaque is predicted by the segmentation decoder. 
Plaque is involved in most lumen narrowings regardless of their functional significance. Therefore, plaque information may play a role in the observed increase in sensitivity and decrease in specificity. 
Including all VAE features, i.e. also background information, led to a slight reduction in sensitivity and the specificity remained the same. This confirms that all relevant information is contained in the VAE features attached to the segmentation decoder, as opposed to the subset of VAE features that describe the background.

The results show that the use of the reference lumen as input improves the pullback prediction (Figure \ref{fig:curves}). This indicates that improvements in the characterization network would also lead to higher performance for pullback prediction. 
The characterization network was trained using X-ray angiography to guide reference annotations, ensuring correct quantification of blooming and non-calcified plaque.
Although the shape of the resulting lumen areas shows good agreement with the reference (Figure \ref{fig:curves}), occasionally an offset is present throughout the artery. This likely relates to blooming, which may be challenging to quantify, as information from X-ray angiography is only used for annotating the reference and hence not available to the network at testing.
The effect that these errors have on the stenosis assessment is mitigated by removing the trend from the lumen area by computing the percentage change to the previous value in the pre-encoder (Section \ref{sec:meth:FFRPred}). 
Another typical error of the characterization network is given by overestimation of the lumen area at bifurcations, which is challenging to define in the reference. To mitigate the impact of bifurcation induced changes in the lumen area on the pullback prediction, we explicitly input the location of bifurcations into the FFR prediction network. 
Finally, our characterization network occasionally fails to identify soft plaque, which can resemble lumen.
As this results in an underestimation of the lumen area decrease at lesions, failure to identify soft plaque is expected to have a significant impact on the performance of FFR prediction. This will be investigated in future work.









One limitation of the study at hand is that the addressed population is heavily diseased, as invasive pullback measurement is only performed for arteries in which a functionally significant stenosis is present. 
However, for the larger group of patients that currently receives only a single FFR measurement, information on the distribution of stenosis may also be of interest. 
Future work may investigate the performance of our method on other target populations.
Furthermore, unlike e.g. Sonck et al. \cite{sonck_clinical_2022}, we did not exclude arteries due to heavy calcifications, bifurcation or ostial lesions, left main disease or severe vessel tortuosity. This likely introduced additional challenges related to the dataset used in this study, potentially playing a role in the overestimation of lower FFR values found in this study. Nevertheless, by not maintaining strict exclusion criteria, our dataset is likely representative of unseen clinical data.

\section{Conclusion}
We proposed a method for non-invasive prediction of the FFR along a coronary artery from CCTA.
To the best of our knowledge, this is the first deep learning method that enables computation on a standard workstation, and without the need for manual correction of the artery lumen segmentation at test time.
The results show promising performance, potentially enabling the distinction between diffuse and focal CAD. This may pave the way towards fast, accurate, and fully automatic prediction of FFR along the artery from CCTA.

\section*{Conflict of Interest Statement}
This work is funded by PIE Medical Imaging B.V.. CC reports receiving institutional research grants from GE Healthcare, Siemens, Insight Lifetech, Coroventis Research, Medis Medical Imaging, Pie Medical Imaging, CathWorks, Boston Scientific, HeartFlow, Abbott Vascular, and consultancy fees from HeartFlow, Abbott Vascular, and Cryotherapeutics. 
II reports institutional research grants from Pie Medical Imaging, the Dutch Technology Foundation with financial support of Pie Medical Imaging and Philips Healthcare and Abbott and HEU grant with participation of Philips Research.

\section*{Acknowledgments}
This work was supported by PIE Medical Imaging B.V..

\bibliographystyle{spiejour}
\bibliography{pullbackReg}

\end{document}